\documentclass[conference]{IEEEtran}
\usepackage{cite}
\usepackage{amsmath,amssymb,amsfonts}
\usepackage{algorithmic}
\usepackage{graphicx}
\usepackage{textcomp}
\usepackage{xcolor}
\usepackage{braket}

\usepackage{comment}
\usepackage{balance}

\usepackage{caption}
\usepackage{subcaption}

\def\BibTeX{{\rm B\kern-.05em{\sc i\kern-.025em b}\kern-.08em
    T\kern-.1667em\lower.7ex\hbox{E}\kern-.125emX}}
\begin{document}

\title{Assessing the Role of Communication in Scalable Multi-Core Quantum Architectures}

\author{\IEEEauthorblockN{Maurizio Palesi}
\IEEEauthorblockA{\textit{University of Catania}\\
Catania, Italy \\
maurizio.palesi@unict.it}
\and
\IEEEauthorblockN{Enrico Russo}
\IEEEauthorblockA{\textit{University of Catania}\\
Catania, Italy \\
enrico.russo@phd.unict.it}
\and
\IEEEauthorblockN{Davide Patti}
\IEEEauthorblockA{\textit{University of Catania}\\
Catania, Italy \\
davide.patti@unict.it}
\and
\IEEEauthorblockN{Giuseppe Ascia}
\IEEEauthorblockA{\textit{University of Catania}\\
Catania, Italy \\
giuseppe.ascia@unict.it}
\and
\IEEEauthorblockN{Vincenzo Catania}
\IEEEauthorblockA{\textit{University of Catania}\\
Catania, Italy \\
vincenzo.catania@unict.it}
}

\maketitle

\begin{abstract}
Multi-core quantum architectures offer a solution to the scalability limitations of traditional monolithic designs. However, dividing the system into multiple chips introduces a critical bottleneck: communication between cores. This paper introduces qcomm, a simulation tool designed to assess the impact of communication on the performance of scalable multi-core quantum architectures. Qcomm allows users to adjust various architectural and physical parameters of the system, and outputs various communication metrics. We use qcomm to perform a preliminary study on how these parameters affect communication performance in a multi-core quantum system.
\end{abstract}


\section{Introduction}
Quantum computing holds immense potential to revolutionize problem-solving, surpassing the capabilities of classical computers. Algorithms like Shor's~\cite{shor1999polynomial} showcase this potential. However, building large-scale quantum computers with thousands of qubits remains a significant challenge, as evidenced by limitations in current models~\cite{arute_nature19}.

Current quantum computers are based on monolithic single-chip architectures that make them difficult to scale beyond a few thousand qubits due to the impracticality of integrating control circuits and per-qubit wiring while maintaining low error rates~\cite{qcbook_2019}. As millions of qubits are predicted to be needed for worthy quantum algorithms~\cite{preskill_2018}, there is a need for scalable solutions able to mitigate crosstalk errors and other issues associated with densely packed qubits on monolithic single-chip quantum processors. A viable solution in this direction is interconnecting multiple quantum chips in a multi-core configuration~\cite{vandersypen_qi17}.

This approach, pioneered by Grover~\cite{grover1997quantum} and Cleve and Buhrman~\cite{cleve_pr97}, allows quantum circuits to be run on separate quantum cores even though they are physically apart. Of course, to take advantage of such a distributed architecture, a communication infrastructure is needed to serve both the quantum and classical communication~\cite{barral2024review}. The quantum communication sub-system allows the transfer of qubit states among cores to perform operations involving qubits mapped onto different cores. The classical communication sub-system allows the transfer of traditional bits mainly used for control and synchronization purposes and overall orchestration of the system.

Several methods have been proposed to build inter-core communication networks for modular quantum computing architectures, such as integrating quantum links to interconnect superconducting chips~\cite{bravyijap22}, ion-shuttling for ion-trapped computers~\cite{kaushal_avsqs20}, photonic networks~\cite{marinelli_arxiv23}, and the teleportation protocol setup for state transfer~\cite{santiago_nanocom21}. In this paper, we focus on a multi-core quantum architecture in which qubits among cores are transferred by means of the teleportation protocol~\cite{gottesman1999quantum} and a Network-on-Chip (NoC) is used to transmit the control bits to support the protocol. We present qcomm~\cite{qcomm_github}, an open-source simulator which allows analysis of the different components that form the overall communication time in multi-core quantum architectures. To the best of our knowledge, this is the first simulator specifically designed to investigate the role played by the classical communication infrastructure in multi-core quantum architectures. By using qcomm, users can explore the interplay between various architectural parameters, physical parameters, and circuit properties to assess the impact of classical communication on the performance of multi-core quantum architectures. This makes it a valuable tool for researchers and designers to explore the communication challenges in these architectures.

The rest of the paper is organized as follows. Sec.~\ref{sec:background} provides the background by introducing the reference system architecture and the basic mechanisms considered for performing operations acting on qubits mapped into different quantum cores. Sec.~\ref{sec:simulator} presents qcomm focusing on its inputs and outputs and on the considered communication model. Sec.~\ref{sec:experiments} presents the results of experiments aimed at assessing the role played by the classical communication system. Finally, Sec.~\ref{sec:conclusion} concludes the work and draws directions for future developments.

\section{Background}
\label{sec:background}

\subsection{System Architecture}
\begin{figure}
    \centering
    \includegraphics[width=0.9\columnwidth]{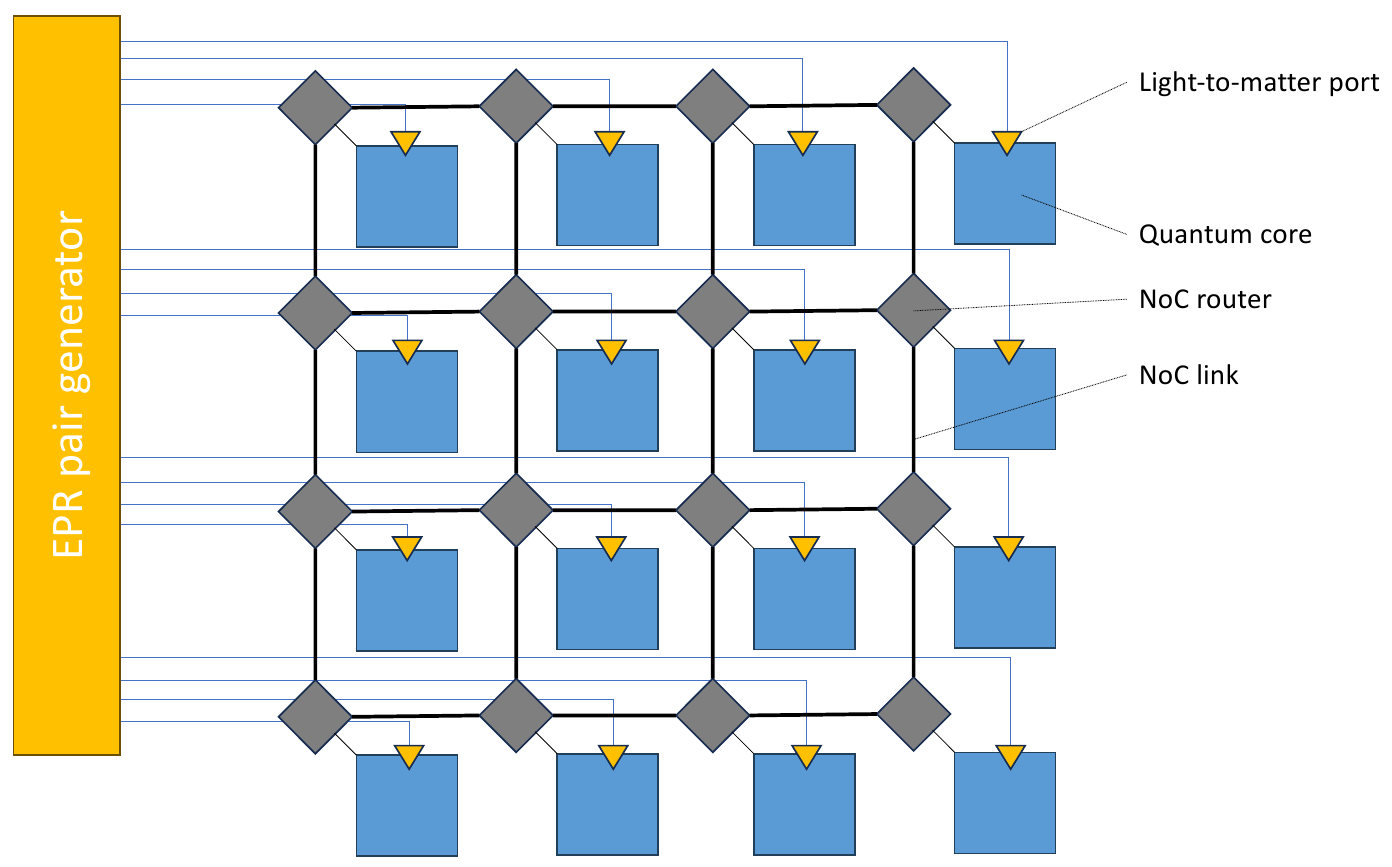}
    \caption{Multi-core quantum architecture.}
    \label{fig:architecture}
\end{figure}

The considered architecture is shown in Fig.~\ref{fig:architecture}. It is formed by a mesh of quantum cores (QCs) interconnected by means of a network-on-chip (NoC). An EPR pair generator allows to generate EPR pairs in parallel~\cite{bravyi_quantum24} that are transmitted to different QCs by means of point-to-point connections. 

\subsection{Teleportation Protocol}
\label{ssec:teleport}
In this paper, we consider intercore communication among QCs achieved through the teleportation protocol, as in~\cite{santiago_nanocom21,escofet_tqc24}. In fact, the qubit is not physically transfered from the source QC to the destination QC. Instead the status of the source qubit is reconstructed into a destination qubit.

\begin{figure}
    \centering
    \includegraphics[width=0.9\columnwidth]{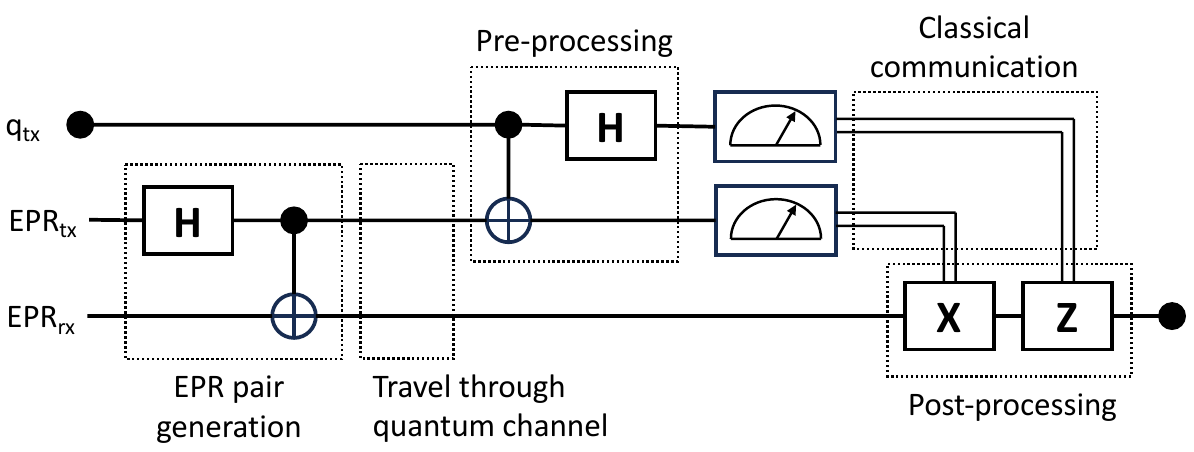}
    \caption{Teleportation protocol.}
    \label{fig:teleportation}
\end{figure}
The teleportation protocol involves five sequential phases, namely, EPR generation, EPR distribution, pre-processing, classical communication, and post-processing that are graphically shown in Fig.~\ref{fig:teleportation}. In EPR generation phases, an entanglement pair of qubits is generated. The two entangled qubits are distributed by means of quantum channels to the two communicating QCs: one to the source QC, namely, $\mathit{EPR}_{tx}$, and one to the destination QC, namely, $\mathit{EPR}_{rx}$. The quantum channels are point-to-point links connecting the EPR pair generator with QCs as shown in Fig.~\ref{fig:architecture} (blue wires). In the pre-processing phase the qubit to be transmitted, namely $q_{tx}$, and $\mathit{EPR}_{tx}$ are pre-processed and measured thus generating two bits of classical information. These two bits are sent to the destination QC by the NoC in the classical communication phase. Finally, in the post-processing phase, the two-bits are used to select one of the  four different quantum operation on $\mathit{EPR}_{rx}$ that results in the transfer of the status of $q_{tx}$ to $\mathit{EPR}_{rx}$. 

\subsection{Remote Gate Operation}
\begin{figure}
    \centering
    \includegraphics[width=0.99\columnwidth]{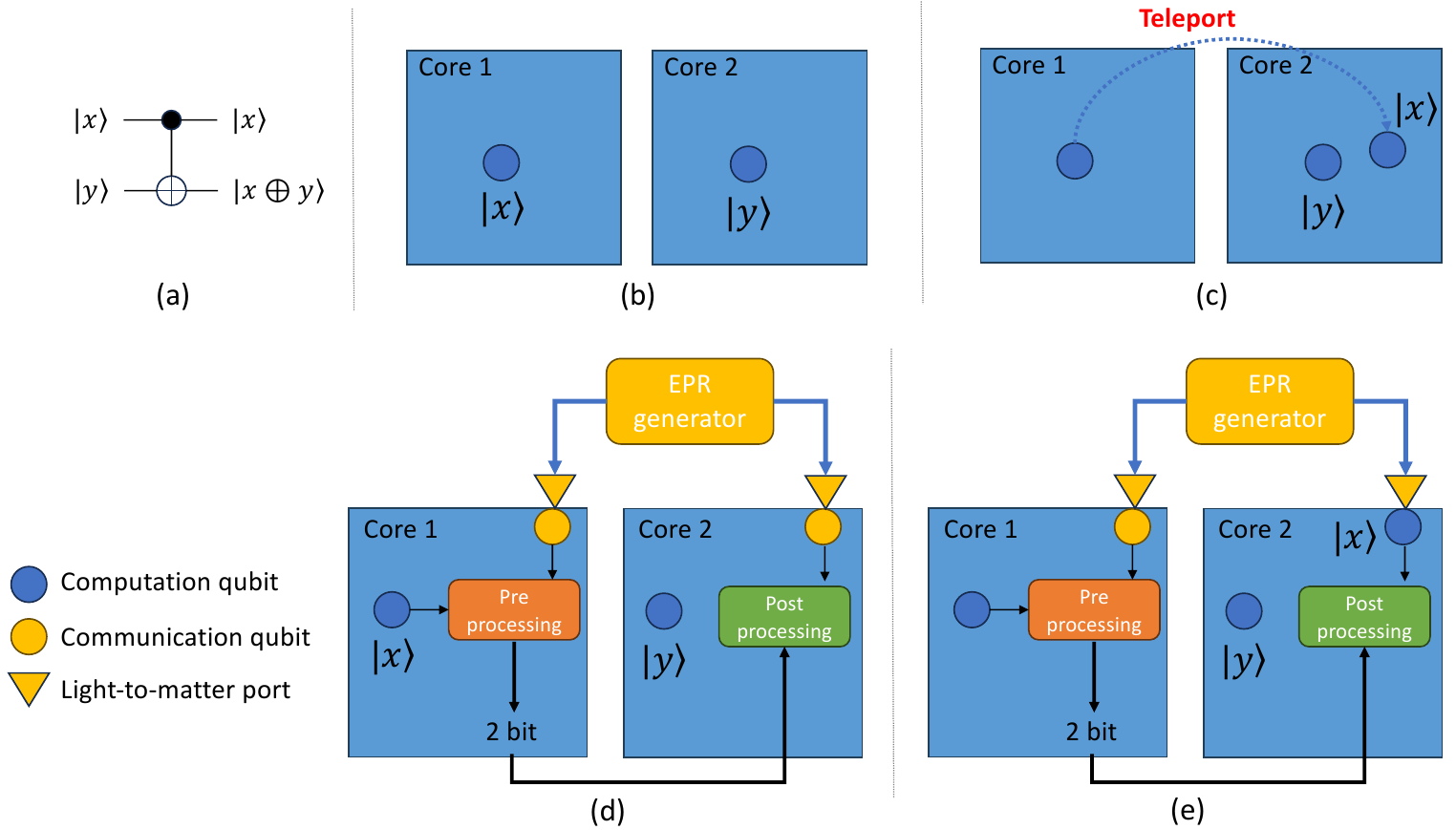}
    \caption{Quantum Teleportation for inter-core CNOT execution: (a) A circuit containing a CNOT operation on qubits $\ket{x}$ and $\ket{y}$. (b) Mapping of the circuit onto a multi-core architecture. Qubits $\ket{x}$ and $\ket{y}$ reside on different cores (core~1 and core~2). (c) Teleportation is used to transfer the state of $\ket{x}$ to core~2, enabling the CNOT operation. (d) Breakdown of the teleportation process: creation of entangled qubits, classical communication, and state recovery using the received classical information. (e) Asuming direct connections between all qubits within the core, the CNOT operation can be performed.}
    \label{fig:teleport-example}
\end{figure}
Circuits being mapped on multi-core quantum architectures often involve gates that must operate on qubits located on different cores~\cite{escofet_tqc24}. Suppose we want to execute a CNOT operation on two qubits, $\ket{x}$ and $\ket{y}$, on such an architecture, as shown in Fig.~\ref{fig:teleport-example}(a). If $\ket{x}$ and $\ket{y}$ are mapped onto different cores (core~1 and core~2, for example, as in Fig.~\ref{fig:teleport-example}(b)), to execute the gate, both qubits must be located on directly connected physical qubits within the same core.

Since we assume all physical qubits within a core are directly connected, either the state of $\ket{x}$ must be transferred to core~2 or the state of $\ket{y}$ must be transferred to core~1. This can be achieved using teleportation, as shown in Fig.~\ref{fig:teleport-example}(c) where the state of $\ket{x}$ is transferred to a qubit in core~2. The different phases involved in the teleportation process are shown in Fig.~\ref{fig:teleport-example}(d). The two entangled qubits generated are transmitted via quantum channels and stored in communication qubits within core~1 and core~2. The pre-processing block in core~1 determines the 2 bits of classical information that are sent to core~2 through the NoC. These two bits are then used by the post-processing block in core~2 to recover the quantum state of $\ket{x}$. However, the original state of qubit $\ket{x}$ in core~1 is lost due to the no-cloning theorem.

At this point, with both $\ket{x}$ and $\ket{y}$ residing in the same core and assuming direct connections between all qubits within the core, the CNOT operation can be performed, Fig.~\ref{fig:teleport-example}(e).

\section{qcomm Simulator}
\label{sec:simulator}
\begin{figure}
    \centering
    \includegraphics[width=0.7\columnwidth]{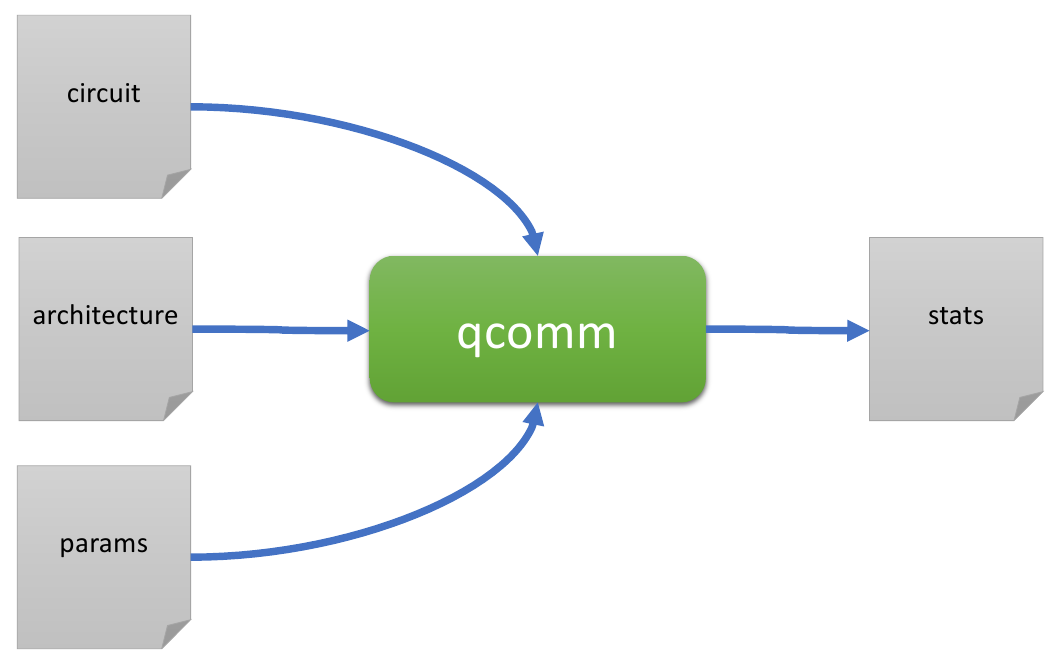}
    \caption{Inputs and output of qcomm.}
    \label{fig:qcomm_io}
\end{figure}
The inputs and output of the simulator are shown in Fig.~\ref{fig:qcomm_io}. Qcomm gets in input the description of the quantum circuit, the description of the architecture, and the physical parameters. It produces in output the execution statistics. Each of the above elements, is presented in the following.

\subsection{Circuit Representation}
\begin{figure}
    \centering
    \includegraphics[width=0.9\columnwidth]{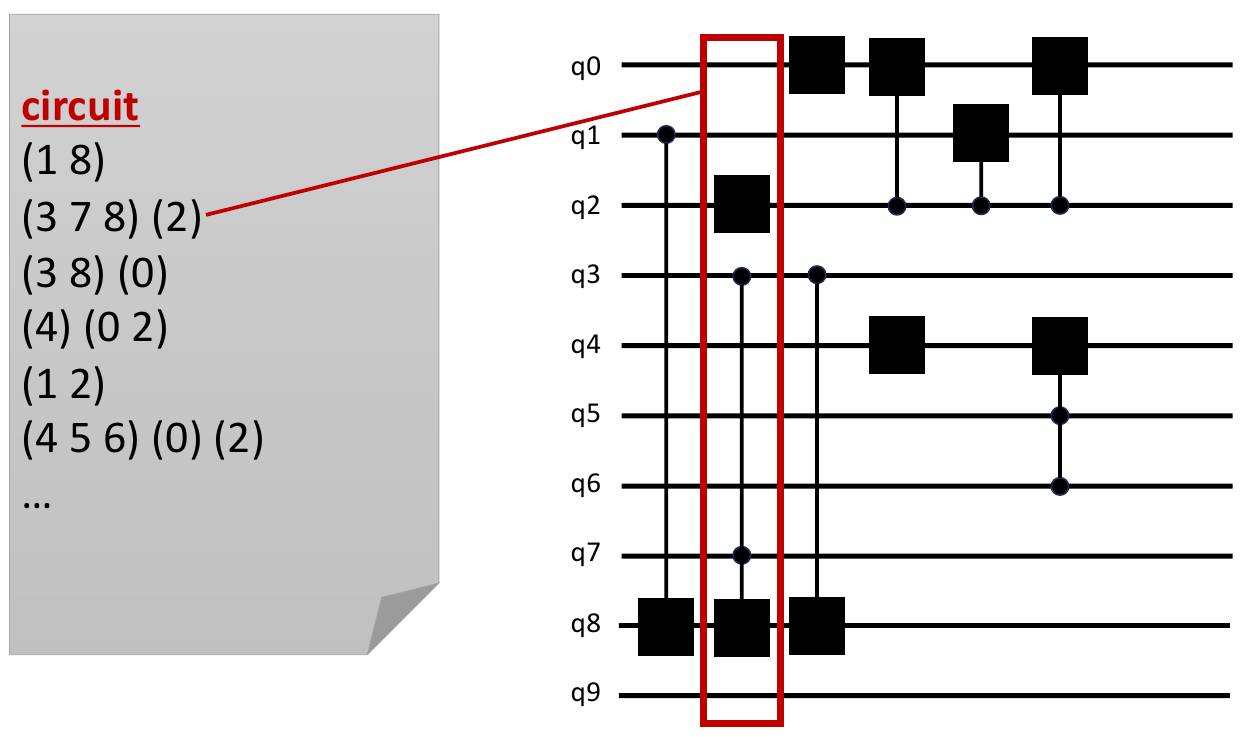}
    \caption{Example of the circuit representation.}
    \label{fig:circuit}
\end{figure}
A quantum circuit is described as a sequence of time slices. Each slice defines a set of gates that can be executed concurrently. Gates within a slice can be executed simultaneously if they share no input qubits. An example circuit description is provided in Fig.~\ref{fig:circuit}. Each line in the circuit file represents a slice, encoded as a list of gates. An $n$-input gate is represented by a tuple containing the qubit indices. For example, in Fig.~\ref{fig:circuit}, the second line represents the second slice. This slice includes a single-input gate involving qubit $q_2$ and a three-input gate involving qubits $q_3$, $q_7$, and $q_8$.

\subsection{Architecture Description}
The architecture file specifies the the resources of the QCs and the characteristics of the NoC describing how these QCs are interconnected. The following parameters are specified in the architecture file:
\begin{itemize}
    \item \texttt{mesh\_x} and \texttt{mesh\_y}: These define the number of QCs in the horizontal and vertical dimensions, respectively. The total number of QCs is the product of \texttt{mesh\_x} and \texttt{mesh\_y}.

    \item \texttt{link\_width}: This defines the width (number of data lines) of the NoC link. It also determines the flit size (a unit of data transferred over the NoC). A communication packet of $n$ bits is divided into $n/$\texttt{link\_width} flits, each with a size of \texttt{link\_width} bits.

    \item \texttt{qubits\_per\_core}: This specifies the total number of physical qubits per QC. Thus, the total number of physical qubits in the system is \texttt{mesh\_x} $\times$ \texttt{mesh\_y} $\times$ \texttt{qubits\_per\_core}.

    \item \texttt{ltm\_ports}: This defines the number of LTM ports per QC. It determines the maximum number of concurrent teleportations that can involve the same QC.
    
    \item \texttt{wireless\_enabled}: This parameter indicates whether the communication system is implemented by a NoC or a wireless NoC (WiNoC)~\cite{deb_jestcs12}.

    \item \texttt{radio\_channels}: This specifies the number of radio channels available in the WiNoC (if enabled by \texttt{wireless\_enabled}).
\end{itemize}

\subsection{Parameters}
\label{ssec:parameters}
The parameters file specifies the physical parameters used in the simulation. These parameters include the gate delay and the delay and data rate figures for the classical communication system (NoC and WiNoC). Specifically, the following parameters are defined:
\begin{itemize}
    \item \texttt{gate\_delay}: Delay of a gate. In the current implementation, all gates are considered to have the same delay regardless of their type or number of inputs.
    
    \item \texttt{hop\_delay}: Delay of a hop in the NoC. This includes both the router delay and the link delay.

    \item \texttt{epr\_delay}: Mean of the EPR pair generation time.

    \item \texttt{dist\_delay}: EPR pair distribution time.

    \item \texttt{pre\_delay} and \texttt{post\_delay}: Delay of the pre-processing and post-processing block for teleportation, respectively.

    \item \texttt{wbit\_rate}: This parameter specifies the WiNoC bit-rate used only if the \texttt{wireless\_enabled} flag in the architecture file is set.

    \item \texttt{token\_pass\_time}: When the communication system is a WiNoC, the medium access control mechanism is based on token passing~\cite{deb_jestcs12}, where only the wireless interface (WI) holding the token can transmit. This parameter represents the time spent by the token to be moved from one WI to another. 
\end{itemize}

\subsection{Statistics}
The output of the simulation is a report containing various execution statistics. These statistics include:
\begin{itemize}
    \item Executed gates: The number of gates in the circuit that have been executed.

    \item Intercore communications: The total number of qubits transferred from one QC to another to perform gates involving qubits located on different QCs.

    \item Throughpunt (peak and average): Peak and average rate of classical communication observed during the execution of the circuit. This refers to the communication overhead (either on the NoC or WiNoC) required to implement the teleportation protocol.
    
    \item Core utilization (average, minimum, and maximum): The average, minimum, and maximum number of qubits in a QC. Initially, the logical qubits of the circuit are uniformly mapped to the QCs. In other words, in a circuit with $n$ qubits, each QC will host $m=n$/(\texttt{mesh\_x} $\times$ \texttt{mesh\_y}) qubits, where $m$ must be less than or equal to \texttt{qubits\_per\_core}. As the simulation progresses, some qubits are redistributed from one QC to another, thus varying the distribution of qubits across the QCs.

    \item Communication time: The portion of the execution time spent on communication. It is further divided into five components representing the time spent for EPR pair generation, EPR pair distribution, pre-processing, classical communication, and post-processing.
    
    \item Computation time: The portion of the execution time spent on computation, i.e., gate execution.

    \item Execution time: Total execution time, which is the sum of the communication time and computation time.
    
    \item Coherence: Based on the experimental results from~\cite{muhonen_nnano14}, the decoherence time of electron spin qubits (without implementing the dynamic decoupling technique for extending qubit lifetime) is $T_2 = 268 \ \mu$s. Assuming an exponential decay of the qubit coherence over time, the qubit coherent state after $t$ $\mu$s is obtained as $e^{-t/T_2}$.
\end{itemize}

\subsection{Communication Model}
The communication time to transfer the state of a qubit from a core $s$ to a core $d$, $\mathit{CT}_{s \rightarrow d}$, is computed as the sum of the times spent in each phase of the teleportation protocol:
\[ \mathit{CT}_{s \rightarrow d} = \mathit{EPRG} + \mathit{EPRD} + \mathit{PRE} + \mathit{CC}_{s \rightarrow d} + \mathit{POST} \]
where $\mathit{EPRG}$ is the EPR generation time, $\mathit{EPRD}$ is the EPR distribution time, $\mathit{PRE}$ is the pre-processing time, $\mathit{CC}_{s \rightarrow d}$ is the classical communication time, and $\mathit{POST}$ is the post-processing time. Please, refer to Sec.~\ref{ssec:teleport} for a refresh of the meaning of the above terms. In our model, apart from the classical communication time which depends on the location of the cores $s$ and $d$ in the NoC, the other terms are considered to be constant.  

We implemented two models for computing $\mathit{CC}_{s \rightarrow d}$ contribution: one for the NoC and the other for the WiNoC.

\subsubsection{NoC Communication Model}
The classical communication time for the NoC is the time spent for sending a packet from a source QC $s$ to a destination QC $d$. It is computed as the product between the hop distance between $s$ and $d$ and the hop time. The hop distance refers to the number of routers and links traversed by the packet assuming minimum routing. The hop time is the sum between router latency and link latency. The link transmission time is the time spent for a packet to traverse the link and depends on the packet size, the link width and clock frequency. 

The packet size in bits is computed as $2 \lceil \log_2 \mathit{nqubits} \rceil + 2$, where the first term represents the bits needed to encode the source and the destination address. The constant 2 accounts for the 2 bits needed to support the teleportation protocol. Because the packet carries source and destination addresses, the number of bits needed to encode them depends on the total number of physical qubits in the system, that is:
\[ \mathit{nqubits} = \texttt{mesh\_x} \times \texttt{mesh\_y} \times \texttt{qubits\_per\_core} \]

All the above information is derived from the parameters specified in the physical parameter file in input to the simulator (Sec.~\ref{ssec:parameters}).

Executing a circuit slice can generate multiple concurrent communication streams. However, our communication model is not a simple zero-contention model. In a zero-contention model, each communication would be independent and have full access to network resources. Instead, our simulator considers the potential overlap of routing paths for these concurrent communications, thus allowing for a more accurate estimation of the classical communication time. When communication streams share the same network link, data packets from these streams are serialized, which increases the overall classical communication time.

\subsubsection{WiNoC Communication Model}
The classical communication time for the WiNoC is the time spent for sending a packet from a source QC $s$ to a destination QC $d$ using the radio channel. In this scenario, there are no wired connections between routers; instead, wireless interfaces (WIs) using radio channels replace them. The packet is serialized and transmitted at a specific data rate.

Only one WI can use the radio medium at a time. A medium access control (MAC) technique based on a token passing mechanism, as described in~\cite{palesi_jlpea15}, is used.  Only the WI holding the token can transmit on the radio channel for a duration sufficient to complete the packet transmission.

The simulator supports the use of multiple radio channels that thus allow the transmission of multiple packets simultaneously, each on a different radio channel.

\subsection{Parallel Teleportation}
Multiple teleportations can be performed in parallel between separate QC pairs. Additionally, if QCs are equipped with multiple LTM ports, parallel teleportations can also be performed between the same QC pair.

\begin{figure}
    \centering
    \includegraphics[width=0.9\columnwidth]{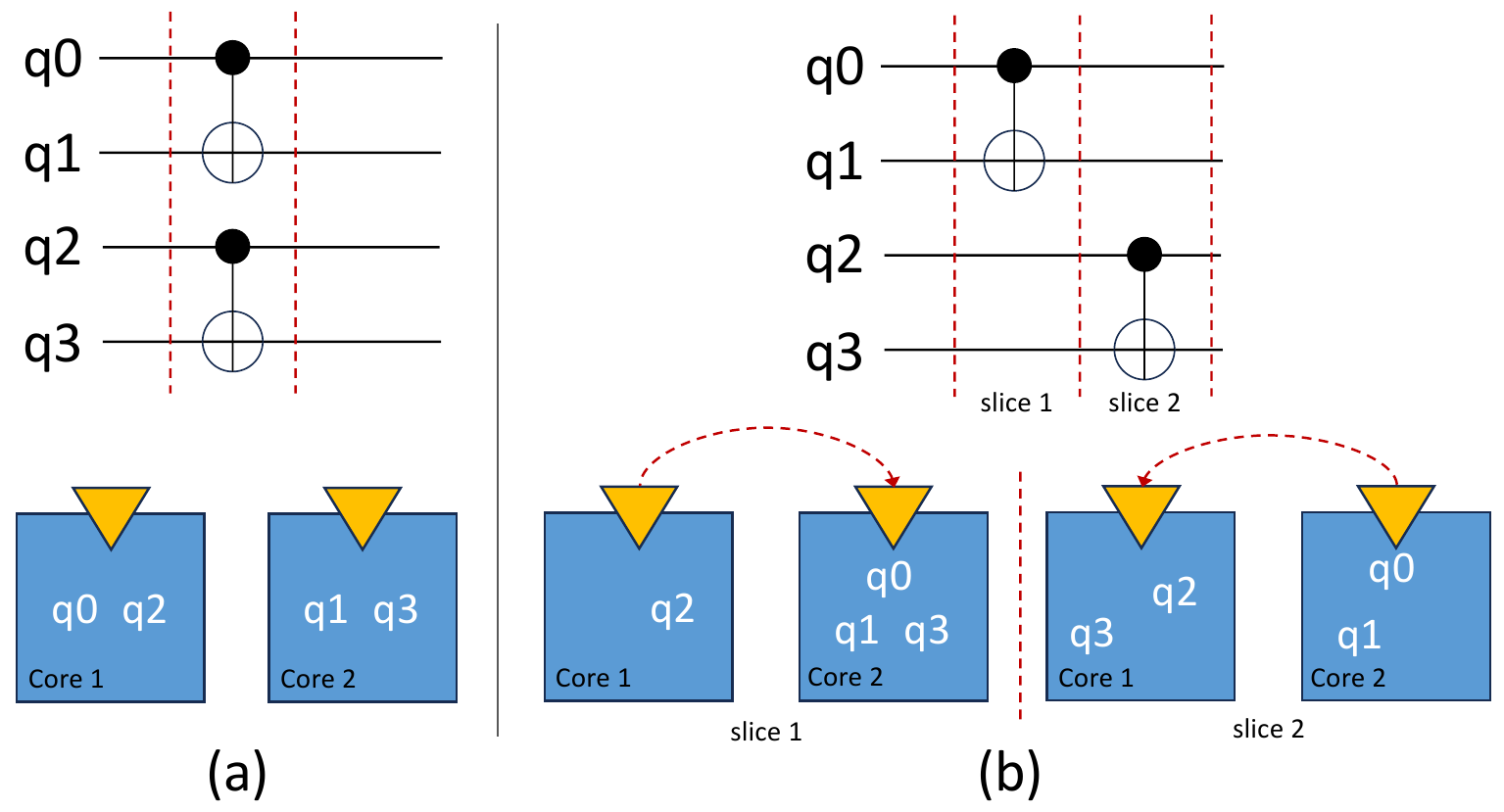}
    \caption{The execution of the two CNOT must be serialized as they involve two teleportation among the same cores that cannot be executed in parallel. The original slice is thus serialized into two slices. In the first slice, the $q_0$ state is transferred to core~2 and the first CNOT is executed. In the second slice, the $q_3$ state is transferred to core~1 and the second CNOT is executed.}
    \label{fig:ltm1}
\end{figure}

Consider the circuit in Fig.~\ref{fig:ltm1}(a), where the two CNOT gates can be executed in parallel. Qubits $q_0$ and $q_2$ are mapped to core~1, while $q_1$ and $q_3$ are mapped to core~2. Since each CNOT gate requires qubits from different cores, neither CNOT can be executed until both qubits are in the same core.

To achieve this, two teleportations are necessary, as shown in Fig.~\ref{fig:ltm1}(b). First, the state of $q_0$ is teleported to core~2, allowing the CNOT between $q_0$ and $q_1$ to proceed. Then, the state of $q_3$ is teleported to core~1, enabling the execution of the second CNOT between $q_2$ and $q_3$.

In essence, even though the CNOT gates can theoretically be executed concurrently because they involve separate qubits, practical limitations arise due to the single LTM port per core.  This single port restricts teleportation to a serial process, preventing true parallel execution of the CNOT gates.

\begin{figure}
    \centering
    \includegraphics[width=0.9\columnwidth]{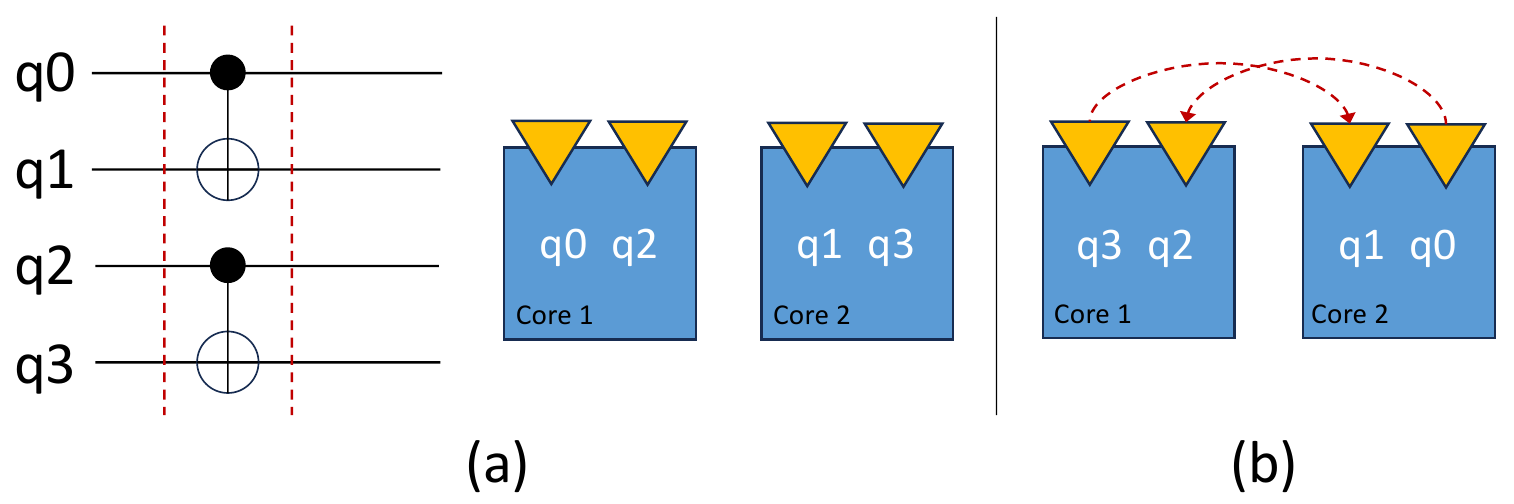}
    \caption{Both CNOT can be executed in parallel as two teleportation, one that transfers the state of $q_0$ to core~2 and the other that transfers the state of $q_3$ to core~0, can be performed in parallel each using a different LTM pair.}
    \label{fig:ltm2}
\end{figure}

Now, consider the scenario in Fig.~\ref{fig:ltm2}(a) where each core has two LTM ports. In this case, both CNOT gates can be executed simultaneously because the transfers of $q_0$ to core~2 and $q_3$ to core~1 can be parallelized, as shown in Fig.~\ref{fig:ltm2}(b). Qcomm factors in the availability of LTM ports when scheduling the execution of gates that involve qubits on different cores. It then splits the time slice based on this availability to optimize scheduling. 

\section{Experiments}
\label{sec:experiments}
In this section, we use qcomm to conduct a series of experiments to assess the role of communication in multi-core quantum architectures. We specifically investigate how communication time varies with different architectural parameters and circuit properties. Tab.~\ref{tab:parameters} reports the parameters used in the experiments, which are based on transmon superconducting qubit technology used in~\cite{morten_arcmp20}.

\begin{table}
    \centering
    \caption{Parameters used for the experiments~\cite{morten_arcmp20}.}
    \label{tab:parameters}
    \begin{tabular}{lc}
        \hline
        Parameter & Value \\
        \hline
        Mean of EPR pair generation time    & $10^3$ ns \\
        EPR pair distribution time          & 0.01 ns \\
        Pre-processing time                 & 390 ns \\
        Classical transfer time (hop delay) & 1 ns \\
        Post-processing time                & 30 ns \\
        \hline
    \end{tabular}
\end{table}

\subsection{Impact of LTM Ports}
In our first experiment, we investigate how communication time changes with the number of LTM ports. We consider a $4 \times 4$ mesh architecture with 16 QCs interconnected by a wired NoC, as shown in Fig.~\ref{fig:architecture}. The NoC utilizes 8-bit links and operates at a clock speed of 1~GHz. Each QC integrates 10 physical qubits, resulting in a total of 160 physical qubits for the entire system.

To analyze communication under varying traffic conditions, we use three randomly generated circuits, each containing 100 qubits and 1,000 gates. These circuits differ in the ratio of 1-input and 2-input gates. The percentage of 2-input gates directly affects communication traffic because these gates require communication when the involved qubits reside on separate QCs. The three circuits have the following ratios: 75\% 1-input gates and 25\% 2-input gates, 50\% of each type, and 25\% 1-input gates and 75\% 2-input gates.

\begin{figure}
    \centering
    \includegraphics[width=0.9\columnwidth]{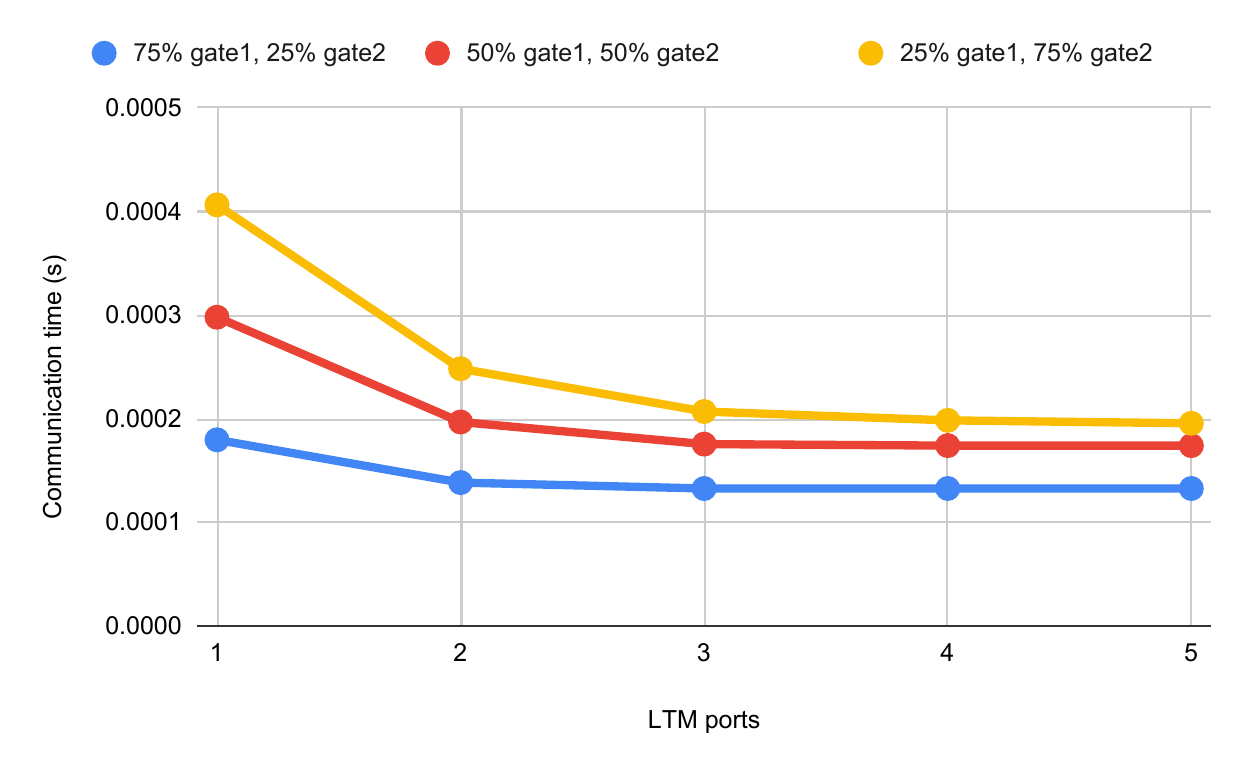}
    \caption{Impact of the number of LTM ports on communication time.}
    \label{fig:ltm_ports}
\end{figure}

Fig.~\ref{fig:ltm_ports} shows the total communication time for different numbers of LTM ports. Circuits with a higher proportion of 2-input gates experience increased communication traffic. This is because these gates necessitate communication when the qubits reside on separate QCs, leading to a rise in communication time. Conversely, as the number of LTM ports increases, communication time decreases due to the increased potential for parallelizing communication between the same QCs. Interestingly, while the decrease in communication time with more LTM ports is an expected outcome, we observe that there's no further improvement beyond 3 LTM ports.

\begin{figure}
    \centering
    \includegraphics[width=0.9\columnwidth]{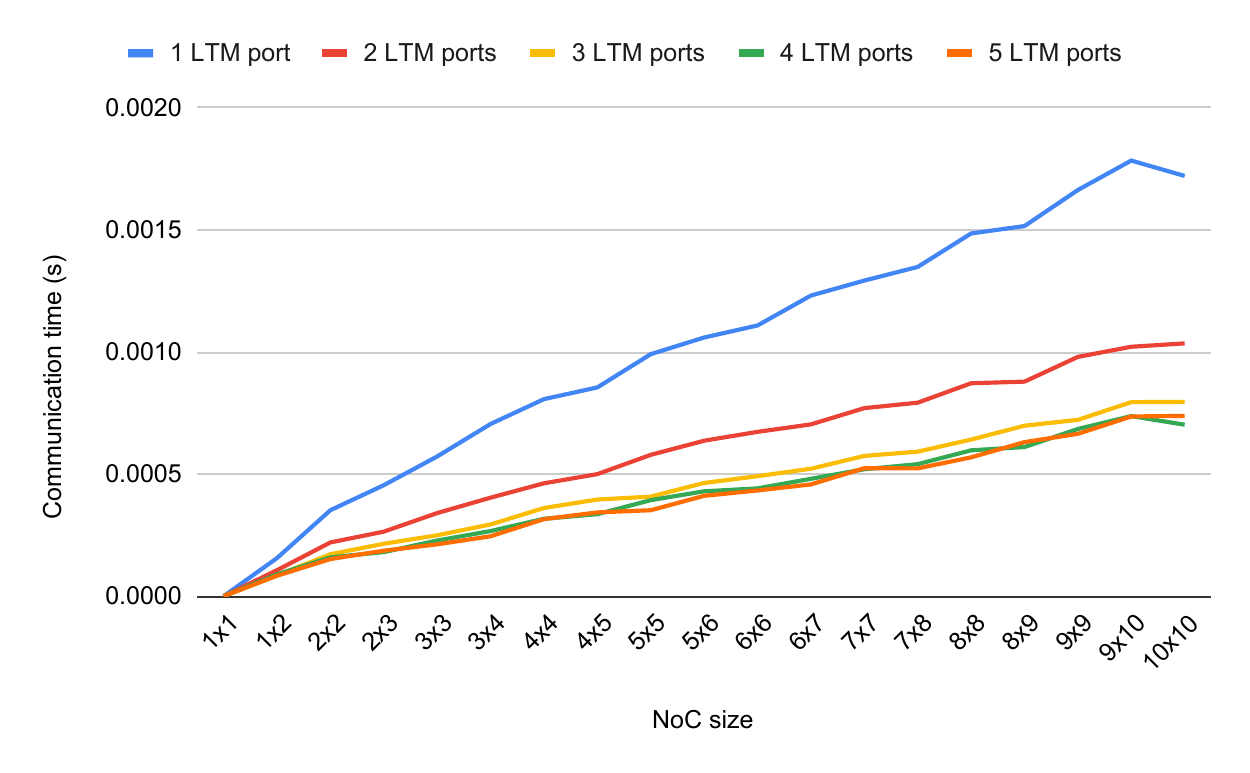}
    \caption{Communication time \emph{vs.} NoC size for different number of LTM ports.}
    \label{fig:ltm_vs_nocsize}
\end{figure}

Our results show that communication improvement plateaus after 3 LTM ports, regardless of the network size (number of QCs). To verify this, we measured communication time as the network size increased (i.e., more QCs). We also proportionally scaled the circuit complexity with the system size. For the single-QC baseline system with 15 physical qubits, the mapped circuit has 10 logical qubits and 100 gates. When simulating an $n$-QC system, we map a circuit with $10 n$ logical qubits and $100 n$ gates. Fig.~\ref{fig:ltm_vs_nocsize} confirms this observation. It shows communication time for different LTM port counts as the number of QCs increases. Consistent with our findings in Fig.~\ref{fig:ltm_ports} for the 16-QC system, communication time saturates when LTM ports exceed 3, irrespective of network size.

\subsection{Communication Time Breakdown}
In the second experiment, we analyze how different NoC architecture characteristics contribute to the overall communication time.

\begin{figure}
    \centering
    \includegraphics[width=0.9\columnwidth]{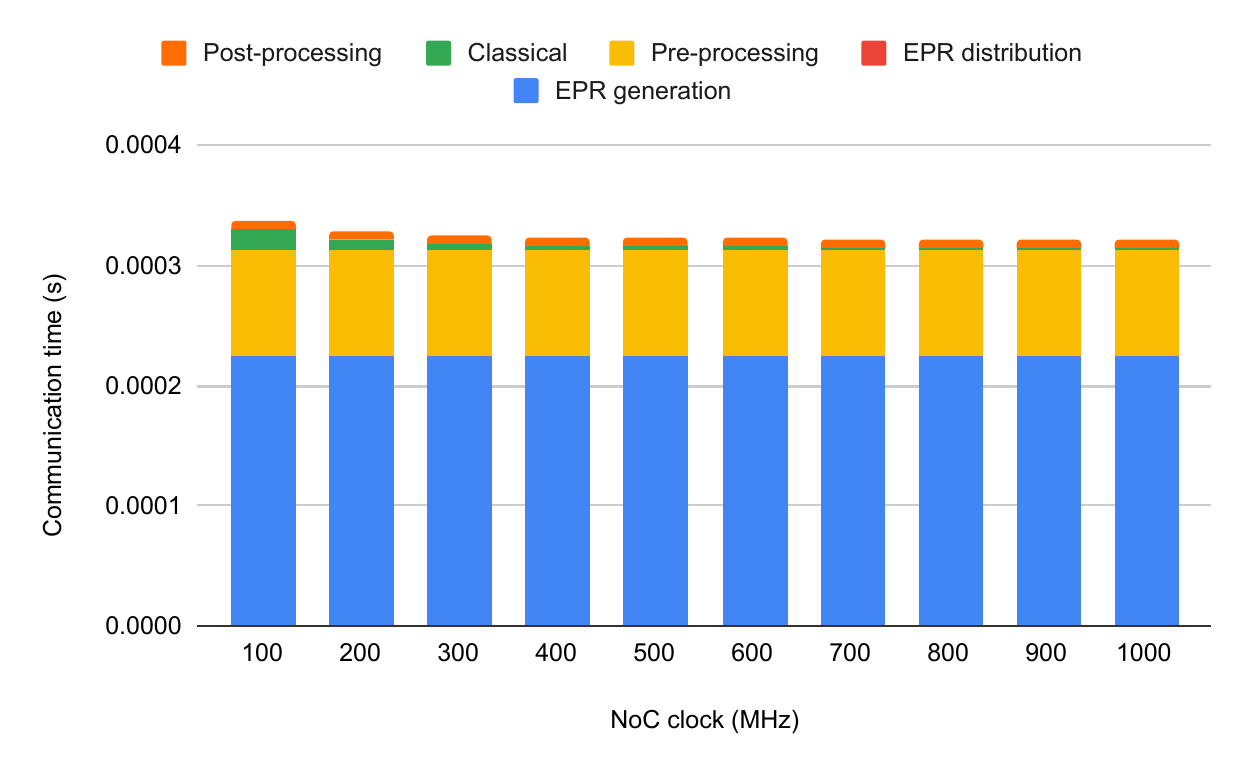}
    \caption{Breakdown of communication time by NoC clock frequency.}
    \label{fig:ct_breakdown_freq}
\end{figure}

We use a $4 \times 4$ NoC (16 QCs) with 8-bit links and 2 LTM ports per QC. The mapped circuit is randomly generated, containing 100 logical qubits and 1,000 2-input gates. Fig.~\ref{fig:ct_breakdown_freq} shows a breakdown of communication time as the NoC clock frequency varies from 100~MHz to 1~GHz. Across all frequencies, the classical communication latency has a minimal impact on the total communication time. The dominant factors are EPR generation time and preprocessing time.

\begin{figure}
    \centering
    \includegraphics[width=0.9\columnwidth]{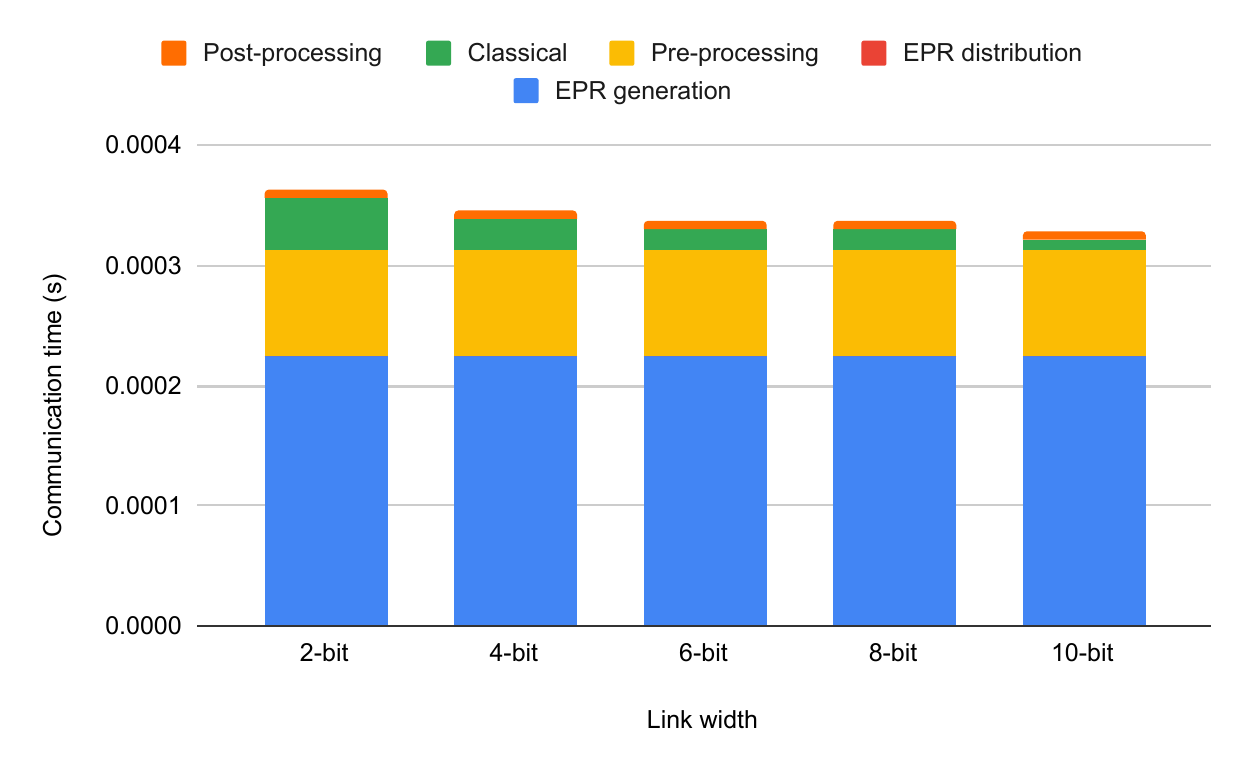}
    \caption{Breakdown of communication time by NoC link width.}
    \label{fig:ct_breakdown_link}
\end{figure}

Fig.~\ref{fig:ct_breakdown_link} analyzes the impact of NoC link width, which varies from 2 to 10 bits. Even with the lowest-performing option (a 2-bit link width), classical communication still has a minimal impact on the total communication time. Here too, EPR generation and preprocessing time remain the dominant factors. In the worst case, classical communication only affects the total communication time by a mere 12\%.

\begin{figure}
    \centering
    \includegraphics[width=0.9\columnwidth]{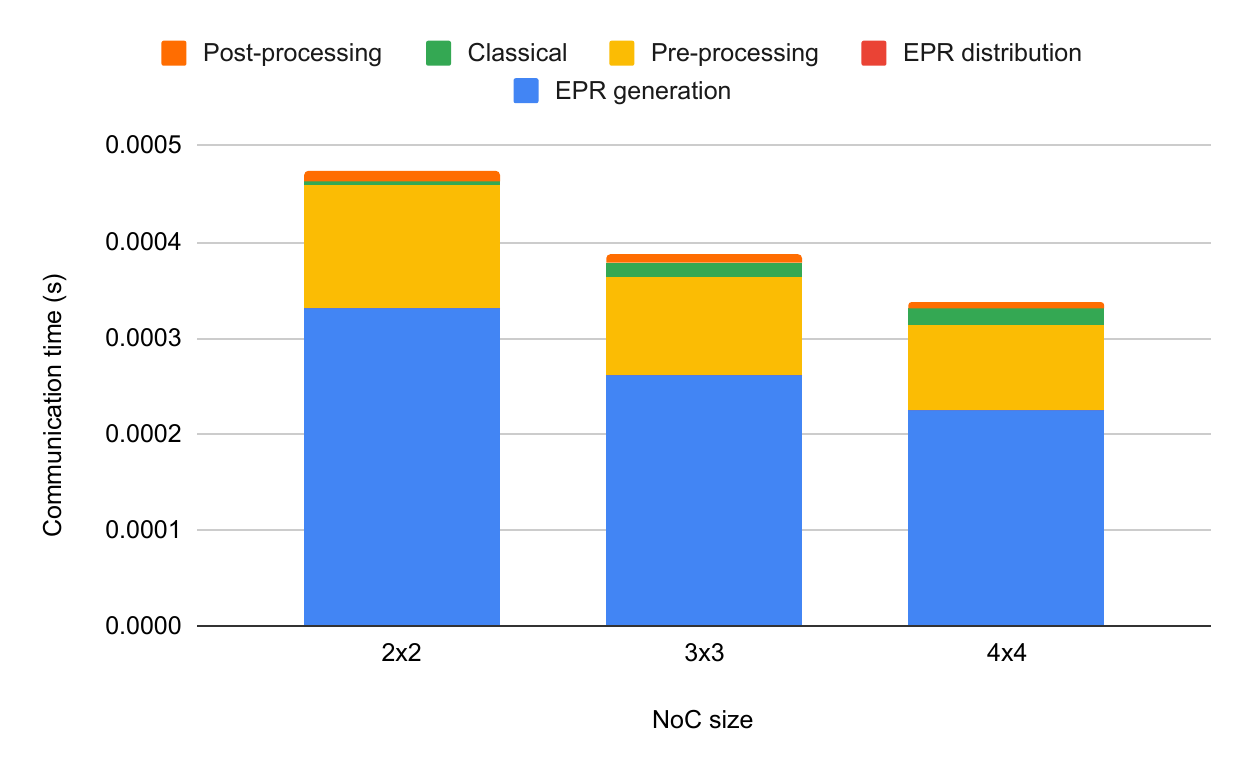}
    \caption{Breakdown of communication time by NoC size.}
    \label{fig:ct_breakdown_nocsize}
\end{figure}

Finally, Fig.~\ref{fig:ct_breakdown_nocsize} presents a breakdown of communication time for different NoC sizes. We use the same mapped circuit for all NoC sizes. As expected, communication time decreases with increasing NoC size due to the potential for greater parallelism. Interestingly, the contribution of classical communication becomes increasingly relevant as the NoC size grows. This is because unlike other factors that influence communication time (e.g., EPR generation, preprocessing), classical communication depends on the distance between communicating nodes. As the NoC size increases, the average communication distance also increases, making classical communication a more significant cost factor.

Across all scenarios examined, classical communication appears to have a minimal impact compared to the time spent on EPR generation and preprocessing.  However, it is important to note that this breakdown is specific to the quantum technology used in our setup (see Tab.~\ref{tab:parameters}). In the next subsection, we will explore how advancements in quantum technology could alter this scenario, potentially making the classical communication system the bottleneck.

\subsection{Previsional Trends}
The previous analysis of the communication system's role is based on current quantum technology parameters, summarized in Tab.~\ref{tab:parameters}. In this scenario, classical communication plays a secondary role compared to other components affecting the total communication time. This subsection explores when classical communication becomes the bottleneck.

\begin{figure}
    \centering
    \begin{subfigure}[b]{0.9\columnwidth}
        \centering
        \includegraphics[width=0.99\textwidth]{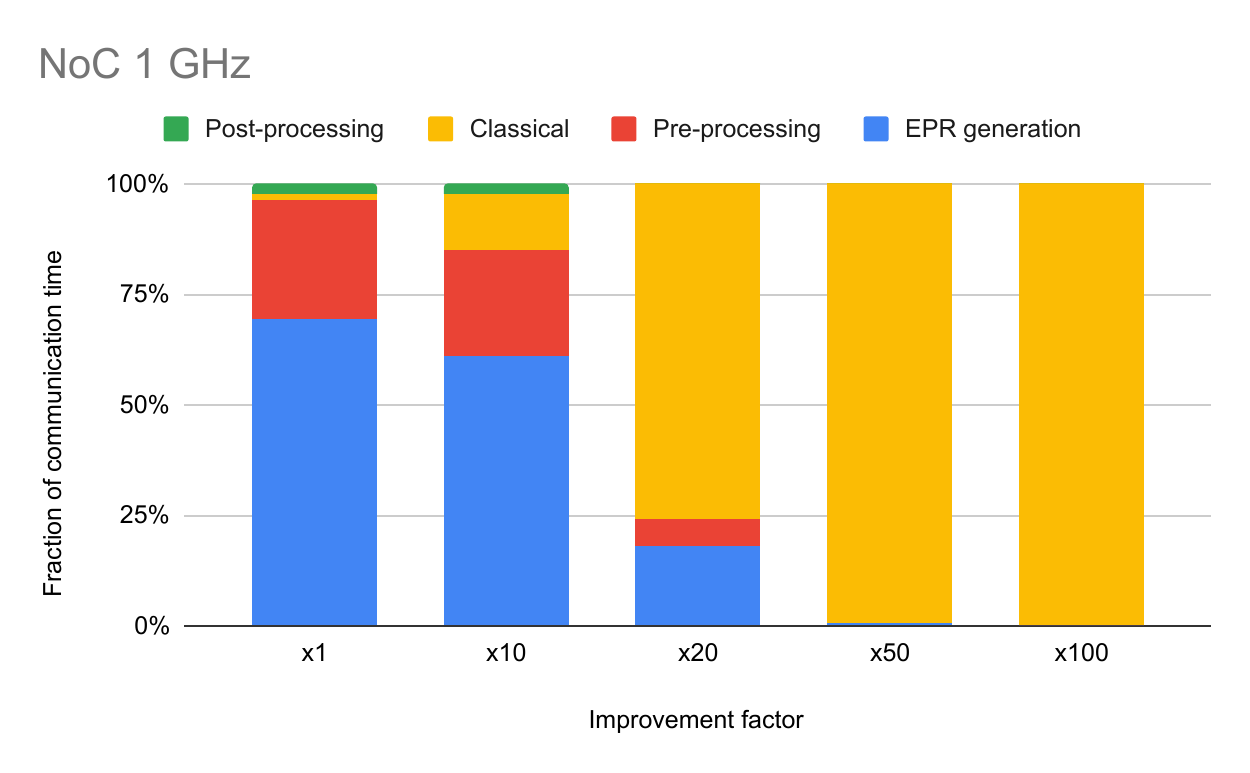}
    \end{subfigure}
    \begin{subfigure}[b]{0.9\columnwidth}
        \centering
        \includegraphics[width=0.99\textwidth]{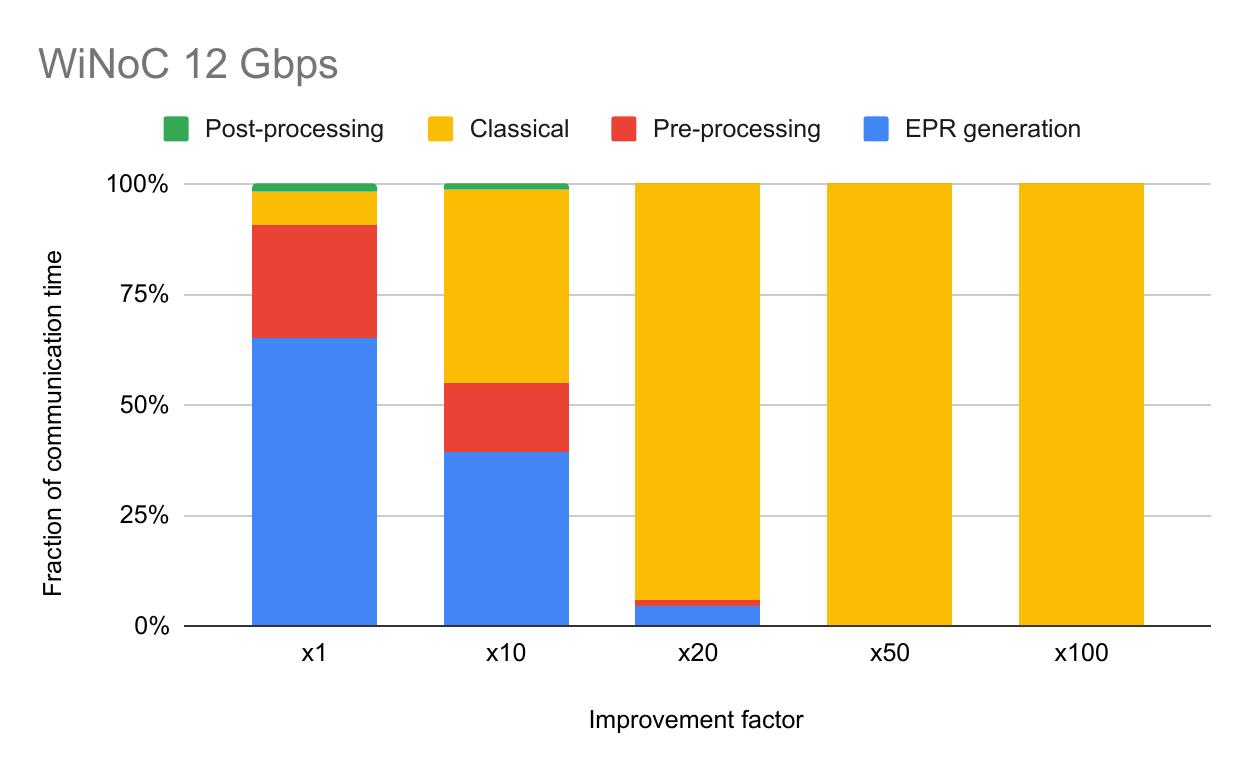}
    \end{subfigure}
    \caption{Fraction of the total communication time accounted by the different contributors when a wired NoC (top) and a wireless NoC (bottom) are used for supporting the classical communication.}
    \label{fig:graph-scaling}
\end{figure}

To investigate this, we progressively scale quantum-related parameters (EPR generation time, pre-processing time, and post-processing time) by the same factor. We aim to identify the point at which classical communication becomes the limiting factor. We consider two communication systems: a traditional wired NoC architecture clocked at 1~GHz and an emerging wireless NoC architecture (WiNoC)~\cite{deb_jestcs12} with a data rate of 12~Gbps. Fig.~\ref{fig:graph-scaling} presents the results, with the improvement factor on the x-axis. As observed in the figure, a 20x improvement in quantum parameters makes the wired NoC the bottleneck, accounting for over 75\% of the total communication time. For the WiNoC, the same improvement factor leads to it accounting for 95\% of the communication time. Even a 10x improvement reveals a substantial contribution from the WiNoC, at 44\% of the total time.

Overall, while classical communication appears marginal compared to other components in the current scenario, this may change as quantum technology advances. Future improvements could make classical communication the bottleneck, highlighting its importance for further research.

\section{Conclusion}
\label{sec:conclusion}
This paper presents qcomm, a novel open-source simulator specifically designed to analyze the impact of classical communication on the performance of multi-core quantum architectures. Qcomm allows users to explore the interplay between various architectural parameters, physical parameters, and circuit properties to assess communication overhead.

The results of our experiments using qcomm show that:
\begin{itemize}
    \item Impact of LTM Ports: Communication time decreases significantly with an increasing number of LTM ports per core, facilitating parallel communication. However, the benefit plateaus after three LTM ports due to limitations in the communication workload generated by the circuits used in our experiments.

    \item Communication Time Breakdown: In the current quantum technology landscape, classical communication has a minimal impact on the total communication time compared to factors like EPR generation and preprocessing.

    \item Future Bottlenecks: We anticipate that advancements in quantum technology will reduce the time spent on tasks like EPR generation and preprocessing, potentially making the classical communication system the performance bottleneck. This emphasizes the need for further research on optimizing classical communication strategies for multi-core quantum architectures.
\end{itemize}

In conclusion, qcomm offers a valuable tool for researchers and designers to explore the communication challenges in multi-core quantum architectures. As classical communication may become a critical factor in future quantum systems, qcomm paves the way for further research on efficient communication protocols and network architectures for high-performance quantum computing.

\section*{Acknowledgement}
Authors gratefully acknowledge funding from the European Commission through HORIZON-EIC-2022-PATHFINDEROPEN01-101099697 (QUADRATURE)

\balance
\bibliographystyle{IEEEtranS}
\bibliography{refs}

\begin{thebibliography}{10}
\providecommand{\url}[1]{#1}
\csname url@samestyle\endcsname
\providecommand{\newblock}{\relax}
\providecommand{\bibinfo}[2]{#2}
\providecommand{\BIBentrySTDinterwordspacing}{\spaceskip=0pt\relax}
\providecommand{\BIBentryALTinterwordstretchfactor}{4}
\providecommand{\BIBentryALTinterwordspacing}{\spaceskip=\fontdimen2\font plus
\BIBentryALTinterwordstretchfactor\fontdimen3\font minus \fontdimen4\font\relax}
\providecommand{\BIBforeignlanguage}[2]{{%
\expandafter\ifx\csname l@#1\endcsname\relax
\typeout{** WARNING: IEEEtranS.bst: No hyphenation pattern has been}%
\typeout{** loaded for the language `#1'. Using the pattern for}%
\typeout{** the default language instead.}%
\else
\language=\csname l@#1\endcsname
\fi
#2}}
\providecommand{\BIBdecl}{\relax}
\BIBdecl

\bibitem{arute_nature19}
F.~Arute, K.~Arya, R.~Babbush, D.~Bacon, J.~C. Bardin, R.~Barends, R.~Biswas, S.~Boixo, F.~G. Brandao, D.~A. Buell \emph{et~al.}, ``Quantum supremacy using a programmable superconducting processor,'' \emph{Nature}, vol. 574, no. 7779, pp. 505--510, 2019.

\bibitem{barral2024review}
D.~Barral, F.~J. Cardama, G.~D{\'\i}az, D.~Fa{\'\i}lde, I.~F. Llovo, M.~M. Juane, J.~V{\'a}zquez-P{\'e}rez, J.~Villasuso, C.~Pi{\~n}eiro, N.~Costas \emph{et~al.}, ``Review of distributed quantum computing. from single qpu to high performance quantum computing,'' \emph{arXiv preprint arXiv:2404.01265}, 2024.

\bibitem{bravyijap22}
S.~Bravyi, O.~Dial, J.~M. Gambetta, D.~Gil, and Z.~Nazario, ``The future of quantum computing with superconducting qubits,'' \emph{Journal of Applied Physics}, vol. 132, no.~16, 2022.

\bibitem{bravyi_quantum24}
S.~Bravyi, Y.~Sharma, M.~Szegedy, and R.~d. Wolf, ``Generating {$k$} {EPR}-pairs from an {$n$}-party resource state,'' \emph{{Quantum}}, vol.~8, p. 1348, May 2024.

\bibitem{cleve_pr97}
R.~Cleve and H.~Buhrman, ``Substituting quantum entanglement for communication,'' \emph{Physical Review A}, vol.~56, no.~2, p. 1201, 1997.

\bibitem{deb_jestcs12}
S.~Deb, A.~Ganguly, P.~P. Pande, B.~Belzer, and D.~Heo, ``Wireless noc as interconnection backbone for multicore chips: Promises and challenges,'' \emph{IEEE Journal on Emerging and Selected Topics in Circuits and Systems}, vol.~2, no.~2, pp. 228--239, 2012.

\bibitem{escofet_tqc24}
P.~Escofet, A.~Ovide, M.~Bandic, L.~Prielinger, H.~van Someren, S.~Feld, E.~Alarc{\'o}n, S.~Abadal, and C.~G. Almud{\'e}ver, ``Revisiting the mapping of quantum circuits: Entering the multi-core era,'' \emph{ACM Transactions on Quantum Computing}, 2024.

\bibitem{gottesman1999quantum}
D.~Gottesman and I.~L. Chuang, ``Quantum teleportation is a universal computational primitive,'' \emph{arXiv preprint quant-ph/9908010}, 1999.

\bibitem{grover1997quantum}
L.~K. Grover, ``Quantum telecomputation,'' \emph{arXiv preprint quant-ph/9704012}, 1997.

\bibitem{qcbook_2019}
E.~Grumbling and M.~Horowitz, \emph{Quantum Computing: Progress and Prospects}.\hskip 1em plus 0.5em minus 0.4em\relax The National Academies Press, 2019.

\bibitem{kaushal_avsqs20}
V.~Kaushal, B.~Lekitsch, A.~Stahl, J.~Hilder, D.~Pijn, C.~Schmiegelow, A.~Bermudez, M.~M{\"u}ller, F.~Schmidt-Kaler, and U.~Poschinger, ``Shuttling-based trapped-ion quantum information processing,'' \emph{AVS Quantum Science}, vol.~2, no.~1, 2020.

\bibitem{morten_arcmp20}
M.~Kjaergaard, M.~E. Schwartz, J.~Braum{\"u}ller, P.~Krantz, J.~I.-J. Wang, S.~Gustavsson, and W.~D. Oliver, ``Superconducting qubits: Current state of play,'' \emph{Annual Review of Condensed Matter Physics}, vol.~11, pp. 369--395, 2020.

\bibitem{marinelli_arxiv23}
B.~Marinelli, J.~Luo, H.~Ren, B.~M. Niedzielski, D.~K. Kim, R.~Das, M.~Schwartz, D.~I. Santiago, and I.~Siddiqi, ``Dynamically reconfigurable photon exchange in a superconducting quantum processor,'' \emph{arXiv preprint arXiv:2303.03507}, 2023.

\bibitem{muhonen_nnano14}
J.~T. Muhonen, J.~P. Dehollain, A.~Laucht, F.~E. Hudson, R.~Kalra, T.~Sekiguchi, K.~M. Itoh, D.~N. Jamieson, J.~C. McCallum, A.~S. Dzurak, and A.~Morello, ``Storing quantum information for 30 seconds in a nanoelectronic device,'' \emph{Nature Nanotechnology}, vol.~9, no.~12, pp. 986--991, Dec 2014.

\bibitem{qcomm_github}
\BIBentryALTinterwordspacing
M.~Palesi, ``qcomm,'' 2024. [Online]. Available: \url{https://github.com/mpalesi/qcomm}
\BIBentrySTDinterwordspacing

\bibitem{palesi_jlpea15}
M.~Palesi, M.~Collotta, A.~Mineo, and V.~Catania, ``An efficient radio access control mechanism for wireless network-on-chip architectures,'' \emph{Journal of Low Power Electronics and Applications}, vol.~5, no.~2, pp. 38--56, 2015.

\bibitem{preskill_2018}
\BIBentryALTinterwordspacing
J.~Preskill, ``Quantum computing in the nisq era and beyond,'' \emph{Quantum}, vol.~2, p.~79, Aug. 2018. [Online]. Available: \url{http://dx.doi.org/10.22331/q-2018-08-06-79}
\BIBentrySTDinterwordspacing

\bibitem{santiago_nanocom21}
S.~Rodrigo, S.~Abadal, C.~G. Almud\'{e}ver, and E.~Alarc\'{o}n, ``Modelling short-range quantum teleportation for scalable multi-core quantum computing architectures,'' in \emph{Proceedings of the Eight Annual ACM International Conference on Nanoscale Computing and Communication}, 2021.

\bibitem{shor1999polynomial}
P.~W. Shor, ``Polynomial-time algorithms for prime factorization and discrete logarithms on a quantum computer,'' \emph{SIAM review}, vol.~41, no.~2, pp. 303--332, 1999.

\bibitem{vandersypen_qi17}
L.~M.~K. Vandersypen, H.~Bluhm, J.~S. Clarke, A.~S. Dzurak, R.~Ishihara, A.~Morello, D.~J. Reilly, L.~R. Schreiber, and M.~Veldhorst, ``Interfacing spin qubits in quantum dots and donors—hot, dense, and coherent,'' \emph{Quantum Information}, vol.~3, no.~34, 2017.

\end{thebibliography}

\end{document}